\documentclass[prd,twocolumn,showpacs,preprintnumbers,amsmath,amssymb,superscriptaddress,floatfix,nofootinbib]{revtex4}

\usepackage{graphicx}
\usepackage{amsmath}
\usepackage{amsfonts}
\usepackage{amssymb}
\usepackage{color}

\newcommand{\Slash}[1]{\ooalign{\hfil/\hfil\crcr$#1$}}

\begin{document}

\title{Role of the $\Lambda^+_c(2940)$ in the $\pi^- p \to D^- D^0 p$ reaction close to threshold}

\author{Ju-Jun Xie} \email{xiejujun@impcas.ac.cn}
\affiliation{Institute of Modern Physics, Chinese Academy of
Sciences, Lanzhou 730000, China} \affiliation{Research Center for
Hadron and CSR Physics, Institute of Modern Physics of CAS and
Lanzhou University, Lanzhou 730000, China} \affiliation{State Key
Laboratory of Theoretical Physics, Institute of Theoretical Physics,
Chinese Academy of Sciences, Beijing 100190, China}

\author{Yu-Bing Dong} \email{dongyb@ihep.ac.cn}
\affiliation{Institute of High Energy Physics, Chinese Academy of
Science, Beijing 100049, China} \affiliation{Theoretical Physics
Center for Science Facilities (TPCSF), CAS, Beijing 100049, China}

\author{Xu Cao} \email{caoxu@impcas.ac.cn}
\affiliation{Institute of Modern Physics, Chinese Academy of
Sciences, Lanzhou 730000, China} \affiliation{Research Center for
Hadron and CSR Physics, Institute of Modern Physics of CAS and
Lanzhou University, Lanzhou 730000, China} \affiliation{State Key
Laboratory of Theoretical Physics, Institute of Theoretical Physics,
Chinese Academy of Sciences, Beijing 100190, China}

\begin{abstract}

We report on a theoretical study of the $\pi^- p \to D^- D^0 p$
reaction near threshold within an effective Lagrangian approach. The
production process is described by $t$-channel $D^{*0}$ meson
exchange, $s$-channel nucleon pole, and $u$-channel
$\Sigma^{++}_{c}$ exchange. In our work, the final $D^0 p$ results
from the ground $\Lambda^+_c(2286)$ state and also dominantly from
the excited $\Lambda^+_c(2940)$ state which is assumed as a $D^{*0}
p$ molecular state with spin parity $J^P = \frac{1}{2}^+$ or
$\frac{1}{2}^-$. We calculate the total cross section of the $\pi^-
p \to D^- D^0 p$ reaction. It is shown that the spin-parity
assignment of $\frac{1}{2}^-$ for $\Lambda^+_c(2940)$ gives a
sizable enhancement for the total cross section in comparison with a
choice of $J^p = \frac{1}{2}^+$. However, our theoretical result of
the total cross section is sensitive to the value of the cutoff
parameter involved in the form factor of the exchanged off-shell
particles. Moreover, we also calculate the second order differential
cross section and find it can be used to determine the parity of the
$\Lambda^+_c(2940)$. It is expected that our model calculations can
be tested by future experiments at J-PARC in Japan.

\end{abstract}
\date{\today}

\pacs{13.75.Cs; 14.20.Dh; 13.30.Eg}
\maketitle

\section{Introduction}{\label{introduction}}

The charmed baryon $\Lambda^+_c(2940)$ was first observed by BABAR
collaboration~\cite{Aubert:2006sp} and later confirmed by the Belle
collaboration~\cite{Abe:2006rz} in 2007. Since its mass
($M_{\Lambda^+_c(2940)} = 2939.3 ^{+1.4}_{-1.5}$ MeV) is close to
the threshold of $D^{*0} p$ ($2945.2$ MeV), and its width is rather
narrow ($\Gamma_{\Lambda^+_c(2940)}=17.5 \pm 5.2 \pm 5.9$ MeV
(\cite{Aubert:2006sp}) the $\Lambda^+_c(2940)$ is explained as a
$D^{*0} p$ hadronic molecular state~\cite{He:2006is}. It was first
found that the molecular structure of the $\Lambda^+_c(2940)$ can
explain the experimental data and that if the $\Lambda^+_c(2940)$ is
a $D^{*0} p$ molecular state it is likely a spin-parity $J^P =
\frac{1}{2}^-$ state~\cite{He:2006is}. In Ref.~\cite{He:2010zq}, it
was pointed out that the $D^* N$ systems may behave as $J^P =
\frac{1}{2}^{\pm}$ and $\frac{3}{2}^{\pm}$ baryon states with a
systematical study of the interaction between $D^*$ and the nucleon.
On the other hand,  the strong two-body decays of the
$\Lambda^+_c(2940)$ have been calculated within the hadronic
molecular approach in Ref.~\cite{Dong:2009tg} and it was concluded
that the $J^P = \frac{1}{2}^+$ assignment for $\Lambda^+_c(2940)$ is
favored. This ansatz for the $\Lambda^+_c(2940)$ has been proved to
be also reasonable for the observed three-body decay modes and
radiative decays~\cite{Dong:2010xv,Dong:2011ys}.

Theoretical studies on the production of the $\Lambda^+_c(2940)$ in
the annihilation process $p \bar{p} \to p D^0 \bar{\Lambda}_c(2286)$
have been carried out in Refs.~\cite{He:2011jp,Dong:2014ksa}, where
the total and differential cross sections of the $p \bar{p} \to p
D^0 \bar{\Lambda}_c(2286)$ reaction were studied. In
Ref.~\cite{He:2011jp},  different assignments ($J^P =
\frac{1}{2}^{\pm}$, $J^P =\frac{3}{2}^{\pm}$, and $J^P =
\frac{5}{2}^{\pm}$) for the $\Lambda^+_c(2940)$ were employed and
the first calculations for the production rates of
$\Lambda^+_c(2940)$ in the $p \bar{p} \to p D^0
\bar{\Lambda}_c(2286)$ and  of $p \bar{p} \to \Sigma^{0,++}_c
\pi^{+,-} \bar{\Lambda}_c(2286)$ processes were performed, however,
the initial state interaction (ISI) and the contribution of $D^*$
meson exchange are not included. While in Ref.~\cite{Dong:2014ksa},
the $\Lambda^+_c(2940)$ was treated as a $J^P = \frac{1}{2}^{+}$ or
as a $\frac{1}{2}^-$ molecular $D^{*0} p$ state, meanwhile, the ISI
as well as the $D$ and $D^*$ mesons exchange are included. Those
predictions of Refs.~\cite{He:2011jp,Dong:2014ksa} could be tested
by future experiments at $\bar{\rm{P}}$ANDA.

In the present work, we try to study this charmed baryon in the
pion-induced reaction related to the experiments at J-PARC where the
expected pion energy will reach over $20$ GeV in the laboratory
frame~\cite{Kim:2014qha}, and therefore, it is sufficient to
reproduce this charmed baryon at J-PARC. It is expected that the
J-PARC in Japan is one of  efficient facilities to study this
charmed baryon. Based on the previous work of
Ref.~\cite{Dong:2014ksa}, and within the assumption that the
$\Lambda^+_c(2940)$ is a $D^* p$ hadronic molecular state, we
investigate the role of $\Lambda^+_c(2940)$ and $\Lambda^+_c(2286)$
in the $\pi^- p \to D^- D^0 p$ reaction with the energy closed to
threshold and with a framework of an effective Lagrangian approach.
Initial interaction between incoming $\pi^-$ and proton is modeled
by an effective Lagrangian which is based on the exchange of the
$D^{*0}$ meson. The $D^0 p$ production proceeds via the
$\Lambda^+_c(2286)$ and $\Lambda^+_c(2940)$ intermediate states. The
total and differential cross sections of the $\pi^- p \to D^- D^0 p$
reaction are calculated with different assignments $J^P =
\frac{1}{2}^+$ and $\frac{1}{2}^-$ for the $\Lambda^+_c(2940)$
resonance for a comparison.

This paper is organized as follows. In sec. II, we will present the
formalism and ingredients necessary for our calculations. Then
numerical results for the total and differential cross sections of
the $\pi^- p \to D^- D^0 p$ reaction and discussions are given in
Sec.~III. A short summary is given in the last section.

\section{Formalism and ingredients}{\label{formalism}}

\begin{figure*}[htbp]
\begin{center}
\includegraphics[scale=0.8]{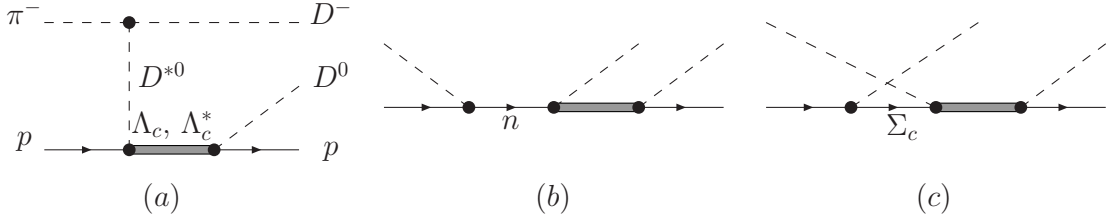} \caption{Feynman
diagrams for the $\pi^- p \to D^- D^0 p$ reaction.} \label{diagram}
\end{center}
\end{figure*}

We study the $\pi^- p \to D^- D^0 p$ reaction within an effective
Lagrangian approach, which has been extensively applied to the study
of scattering
processes~\cite{Tsushima:1998jz,Sibirtsev:2005mv,Liu:2006tf,
Maxwell:2012zz,Liu:2012kh,Xie:2013wfa,Liu:2011sw,Liu:2012ge,Lu:2013jva,Lu:2014rla,Xie:2007qt,Xie:2013mua,Xie:2010yk}
for the production of light baryon states. The basic tree level
Feynman diagrams for the $\pi^- p \to D^- D^0 p$  reaction are
depicted in Fig.~\ref{diagram}. It is assumed that the $D^0 p$ final
states are produced by the decay of the intermediate
$\Lambda^+_c(2286)$ ($\equiv \Lambda_c$) and $\Lambda^+_c(2940)$
($\equiv \Lambda^*_c$) states as the result of the $D^{*0}$ meson
exchanges [Fig.~\ref{diagram} (a)]. Moreover, the contributions
including the $s$-channel nucleon pole [Fig.~\ref{diagram} (b)], and
$u$-channel $\Sigma^{++}_{c}$ [Fig.~\ref{diagram} (c)] are also
considered.

To compute the amplitudes of these diagrams shown in
Fig.~\ref{diagram}, the effective Lagrangian densities for the
relevant interaction vertexes are needed. We use the commonly
employed Lagrangian densities for $D^* D \pi$, $\pi N N$, $D N
\Sigma_c$, $\Lambda_c p D$, $\Lambda_c p D^*$, and $\Lambda_c \pi
\Sigma_c$ as
follows~\cite{Dong:2010xv,Dong:2011ys,Dong:2014ksa,Liang:2004sd,Chen:2011xk,Liu:2001ce}:

\begin{eqnarray}
{\cal L}_{D^* D \pi} &=& g_{D^* D \pi} D^{*}_{\mu}\vec \tau \cdot (D
\partial^{\mu} \vec \pi - \partial^{\mu} D \vec \pi ), \label{ldstardpi} \\
{\cal L}_{\pi N N}  &=& - i g_{\pi N N} \bar{N} \gamma_5 \vec\tau \cdot \vec\pi N, \label{lpinn} \\
{\cal L}_{D N \Sigma_{c}} &=& - i g_{D N \Sigma_{c}} \bar{N}
\gamma_5 D \Sigma_{c} + {\rm H.c.}, \label{ldnsigmac} \\
{\cal L}_{\Lambda_c p D}  &=& i g_{\Lambda_c p
D} \bar{\Lambda}_c \gamma_5 p D^0 + {\rm H.c.}, \label{llcpd} \\
{\cal L}_{\Lambda_c p D^*}  &=& g_{\Lambda_c p D^*} \bar{\Lambda}_c
\gamma^{\mu} p D^{*0}_{\mu} + {\rm H.c.}, \label{llcpdstar} \\
{\cal L}_{\Lambda_c \pi \Sigma_c}  &=& i g_{\Lambda_c \pi \Sigma_c}
\bar{\Lambda}_c \gamma_5  \vec\pi \cdot \vec\Sigma_c + {\rm H.c.}.
\label{llcpd}
\end{eqnarray}

The coupling constants $g_{D N \Sigma_c} = -2.69$, $g_{\Lambda_c p
D} = -13.98$, $g_{\Lambda_c p D^*} = -5.20$, and $g_{\Lambda_c \pi
\Sigma_c} = 9.32$ are determined from $SU(4)$ invariant
Lagrangians~\cite{Dong:2010xv,Liu:2001ce,Okubo:1975sc} in terms of
$g_{\pi NN} = 13.45$ and $g_{\rho NN} = 6$. Besides, the coupling
constant $g_{D^* D \pi}$ can be evaluated from the partial decay
width of $D^* \to D \pi$,
\begin{eqnarray}
\Gamma[D^{*0} \to D^0 \pi^0] &=& \frac{g^2_{D^* D \pi}}{24 \pi}
\frac{|\vec{p}_{\pi}|^3}{M_{D^{*0}}^2} ,
\end{eqnarray}
with $\vec{p}_{\pi}$ the three-momentum of $\pi^0$ in the $D^{*0}$
rest frame. Unfortunately, only an upper bound for this decay rate
is known at present~\cite{Agashe:2014kda}. Here, we take
$\Gamma_{D^{*0}}$ as the same as the total decay width of $D^{*+}$,
which is $\Gamma_{D^{*0}} = \Gamma_{D^{*+}} = 83.4$
keV~\cite{Agashe:2014kda}. With a value of 0.62 (different from 2/3
due to the breaking of isospin symmetry)  for the $D^{*0} \to D^0
\pi^0$ branching ratio, we get $g_{D^* D \pi} =14.1$.~\footnote{The
value we obtained here is in agreement with the value $12.5 \pm 1.0$
that was obtained with QCD sum rules in Ref.~\cite{Belyaev:1994zk}.}

For the $\Lambda^+_c(2940) p D$ and $\Lambda^+_c(2940) p D^*$
couplings, we take the interaction Lagrangian densities as used in
Ref.~\cite{Dong:2014ksa},
\begin{eqnarray}
{\cal L}^{\frac{1}{2}^+}_{\Lambda^*_c p D}  &=& i g_{\Lambda^*_c p
D} \bar{\Lambda}^*_c \gamma_5 p D^0 + {\rm H.c.}, \label{llcpdplus} \\
{\cal L}^{\frac{1}{2}^+}_{\Lambda^*_c p D^*}  &=& g_{\Lambda^*_c p
D^*} \bar{\Lambda}^*_c \gamma^{\mu} p D^{*0}_{\mu} + {\rm H.c.},
\label{llcpdstarplus}
\end{eqnarray}
for the assignment $J^P=\frac{1}{2}^+$ for $\Lambda^+_c(2940)$, and
\begin{eqnarray}
{\cal L}^{\frac{1}{2}^-}_{\Lambda^*_c p D}  &=& f_{\Lambda^*_c p
D} \bar{\Lambda}^*_c p D^0 + {\rm H.c.}, \label{llcpdminus} \\
{\cal L}^{\frac{1}{2}^-}_{\Lambda^*_c p D^*}  &=& - f_{\Lambda^*_c p
D^*} \bar{\Lambda}^*_c  \gamma_5 \gamma^{\mu} p D^{*0}_{\mu} + {\rm
H.c.}, \label{llcpdstarminus}
\end{eqnarray}
for the assignment $J^P=\frac{1}{2}^-$ for $\Lambda^+_c(2940)$.

The couplings $g_{\Lambda^*_c p D^*}$, $g_{\Lambda^*_c p D}$ and
$f_{\Lambda^*_c p D^*}$, $f_{\Lambda^*_c p D}$ in the above
Lagrangians have been evaluated in
Refs.~\cite{Dong:2009tg,Dong:2010xv} using the hadronic molecular
approach with $g_{\Lambda^*_c p D^*} = 6.64$, $g_{\Lambda^*_c p D} =
-0.54$, $f_{\Lambda^*_c p D^*} = 3.75$, and $f_{\Lambda^*_c p D} =
-0.97$. In Ref.~\cite{Dong:2014ksa}, these values are also employed in
the calculation of the annihilation process of $\bar{p} p \to p D^0
\bar{\Lambda}_c (2940)$.

Since the hadrons are not pointlike particles, the form factors are
also needed. For the exchanged $D^{*0}$ meson, we adopt the monopole
form factor following that used in
Refs.~\cite{He:2011jp,Dong:2014ksa,Oh:2007jd,Haidenbauer:2009ad},
\begin{equation}
F_{D^*}(q^{2}_{ex},M_{ex}) = \frac{\Lambda_{D^*}^2 -
M^2_{D^*}}{\Lambda_{D^*}^2 - q^2_{D^*}},
\end{equation}
and for the exchanged baryons, we take the form factor employed in Refs.~\cite{Feuster:1997pq,Shklyar:2005xg},
\begin{equation}
F_B(q^{2}_{ex},M_{ex}) = \frac{\Lambda_B^4}{\Lambda_B^4 +
(q^{2}_{ex}-M^2_{ex})^2}.
\end{equation}
Here the $q_{ex}$ and $M_{ex}$ are the four-momentum and the mass of
the exchanged hadron, respectively. In our present calculation, we
use the cutoff parameters $\Lambda=\Lambda_{D^*} = \Lambda_{N} =
\Lambda_{\Sigma_{c}} = \Lambda_{\Lambda^*_c} = 3$
GeV~\footnote{Actually, the values of the cutoff parameters can be
directly related to the hadron size. Since the question of hadron
size is still very open, we have to adjust those cutoff parameters
to fit the related experimental data. When choosing $\Lambda = 3$
GeV, we follow the argument given in
Refs.~\cite{Dong:2014ksa,Haidenbauer:2009ad}, where such a value was
employed.} for minimizing the free parameters.

The propagator for the exchanged $D^{*0}$ meson used in our
calculation is
\begin{equation}
G^{\mu \nu}_{D^*}(q_{D^*})=\frac{-i(g^{\mu \nu} -
q_{D^*}^{\mu}q_{D^*}^{\nu}/M^2_{D^*})}{q^2_{D^*} - M_{D^*}^2}.
\end{equation}
For the propagator of the spin-1/2 baryon, we use
\begin{equation}
G_{\frac{1}{2}}(q) = \frac{i(\Slash q + M)}{q^2 - M^2+ iM \Gamma} ,
\end{equation}
where $q$ and $M$ stand for the four-momentum and the mass of the
intermediate nucleon pole, $\Sigma_c$ baryon, $\Lambda_c(2286)$
state, and $\Lambda_c(2940)$ resonance, respectively. Since $q^2 <
0$ for $u$-channel $\Sigma_c$ exchange, we take $\Gamma = 0$ for
$\Sigma_c$ and also for the nucleon pole and $\Lambda_c(2286)$
state, while for the $\Lambda_c(2940)$ resonance, we take $\Gamma =
17$ MeV~\cite{Agashe:2014kda}.

From the above effective Lagrangian densities, the scattering
amplitudes for the $\pi^- p \to D^- D^0 p$ reaction can be obtained
straightforwardly. For example, the amplitudes due to the $D^{*0}$
exchange can be written as
\begin{eqnarray}
{\cal M}^{\frac{1}{2}^{\pm}}_{a} & = & \frac{i
g^{\frac{1}{2}^{\pm}}_a}{(q^2 - M^2_{\Lambda'_c} + i
M_{\Lambda'_c} \Gamma_{\Lambda'_c}) (t - M^2_{D^*})} \\ \nonumber
&\times &\bar{u}(p_5,s_f)  (\Slash q \mp M_{\Lambda'_c})(\Slash p_1 - \frac{p_1 \cdot k_t
\Slash k_t}{M^2_{D^*}}) \gamma_5 u(p_2,s_i),
\end{eqnarray}
for Fig.~\ref{diagram} (a), and
\begin{eqnarray}
{\cal M}^{\frac{1}{2}^{\pm}}_b & = & \frac{ \sqrt{2}
g^{\frac{1}{2}^{\pm}}_b}{(q^2 - M^2_{\Lambda'_c} + i
M_{\Lambda'_c} \Gamma_{\Lambda'_c}) (s - m^2_n)} \\ \nonumber
&\times &\bar{u}(p_5,s_f) (\Slash q \mp M_{\Lambda'_c})(\Slash k_s + m_n) \gamma_5 u(p_2,s_i),  \\
{\cal M}^{\frac{1}{2}^{\pm}}_c & = & \frac{
g^{\frac{1}{2}^{\pm}}_c}{(q^2 - M^2_{\Lambda'_c} + i
M_{\Lambda'_c} \Gamma_{\Lambda'_c}) (u - M^2_{\Sigma_c})} \\ \nonumber
&\times&\bar{u}(p_5,s_f)  (\Slash q \mp M_{\Lambda'_c})(\Slash k_u + M_{\Sigma_c}) \gamma_5
u(p_2,s_i), ,
\end{eqnarray}
for Figs.~\ref{diagram} (b) and \ref{diagram} (c), respectively.
Here $p_1$, $p_2$, $p_3$, $p_4$, and $p_5$ are the four-momenta of
the $\pi^-$, initial proton, $D^-$, $D^0$, and final proton,
respectively; $s_i$ and $s_f$ are the spin projections of the
initial and final protons, respectively; $k_{t} = p_1 - p_3$, $k_s =
p_1 + p_2$, and $k_u = p_2 - p_3$ are the four-momenta for the
exchanged $D^{*0}$ meson in $t$ channel, nucleon pole in $s$
channel, and $\Sigma_c$ in $u$ channel, respectively. In the above
equations, $s = k^2_s$, $t = k^2_t$, and $u = k^2_u$ indicate the
Mandelstam variables. The couplings $g^{\frac{1}{2}^{\pm}}_{a,b,c}$
are defined as~\footnote{Since the spin parity of $\Lambda_c(2286)$
is $J^P = 1/2^+$, we replace $g^{\frac{1}{2}^+}_i$ ($i = a, b, c$)
by $g_i$ ($g_{a} = g_{D^* D \pi} g_{\Lambda_c p D^*} g_{\Lambda_c p
D}$, $g_{b} = - g_{\pi NN} g^2_{\Lambda_c p D}$, and $g_{c} = -g_{D
N \Sigma_c} g_{\Lambda_c \pi \Sigma_c} g_{\Lambda_c p D}$) then we
can get the scattering amplitude for the case of the
$\Lambda_c(2286)$ state.}
\begin{eqnarray}
g^{\frac{1}{2}^{+}}_{a} &=& g_{D^* D \pi} g_{\Lambda^*_c p D^*}
g_{\Lambda^*_c p D} , \\
g^{\frac{1}{2}^{-}}_{a} &=& g_{D^* D \pi} f_{\Lambda^*_c p D^*}
f_{\Lambda^*_c p D} , \\
g^{\frac{1}{2}^{+}}_{b} &=& - g_{\pi NN} g^2_{\Lambda^*_c p D} , \\
g^{\frac{1}{2}^{-}}_{b} &=& g_{\pi NN} f^2_{\Lambda^*_c p D} , \\
g^{\frac{1}{2}^{+}}_{c} &=& -g_{D N \Sigma_c} g_{\Lambda^*_c \pi
\Sigma_c} g_{\Lambda^*_c p D} , \\
g^{\frac{1}{2}^{-}}_{c} &=& -g_{D N \Sigma_c} f_{\Lambda^*_c \pi
\Sigma_c} f_{\Lambda^*_c p D} .
\end{eqnarray}

Then the calculations of the differential and total cross sections
for the $\pi^- p \to D^- D^0 p$ reaction are,
\begin{eqnarray}
&& d\sigma (\pi^- p \to D^- D^0 p) = \frac{m_p}{2 \sqrt{(p_1 \cdot
p_2)^2 - m^2_{\pi^-} m^2_p}} \sum_{s_i, s_f} |{\cal M}|^2   \nonumber\\
&& \times \frac{d^{3} p_3}{2 E_3} \frac{d^{3} p_4}{2 E_4} \frac{m_p
d^{3} p_5}{E_5} \delta^4(p_1+p_2-p_3-p_4-p_5), \label{eqcs}
\end{eqnarray}
where $E_3$, $E_4$, and $E_5$ stand for the energy of the $D^-$,
$D^0$, and final proton, respectively.

\section{Numerical results and discussions}

In this section we show our theoretical numerical results for the
total and differential cross sections of the $\pi^- p \to D^- D^0 p$
reaction near the reaction threshold.

\subsection{Total cross sections}

\begin{figure}[htbp]
\begin{center}
\includegraphics[scale=0.45]{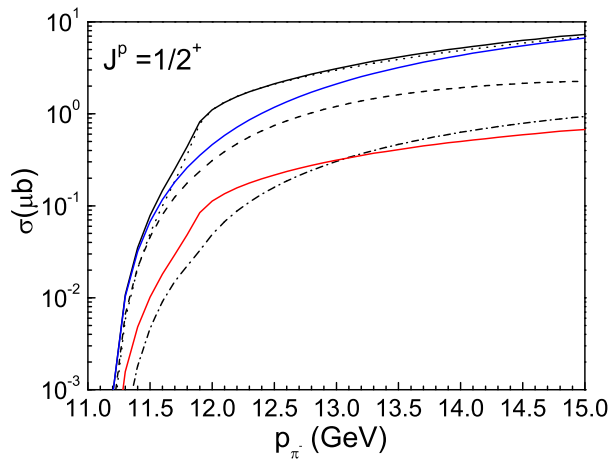}
\caption{(Color online) Total cross sections for the $\pi^- p \to
D^- D^0 p$ reaction as a function of the beam momentum $p_{\pi^-}$
for $J^P = \frac{1}{2}^+$ of the $\Lambda^+_c(2940)$. The dashed,
dotted, and dash-dotted curves stand for the contributions from the
$s$ channel, $t$ channel, and $u$ channel, respectively. Their total
contribution is shown by the solid line. The blue line stands for
the contributions from the ground $\Lambda^+_c(2286)$ state.}
\label{Fig:tcsplus}
\end{center}
\end{figure}

With the formalism and ingredients given above, the total cross
section versus the beam momentum $p_{\pi^-}$ for the $\pi^- p \to
D^- D^0 p$ reaction is calculated by using a Monte Carlo
multiparticle phase space integration program. The theoretical
numerical results obtained with cutoff $\Lambda = 3$ GeV for the
total cross section for $J^P = \frac{1}{2}^+$ of the
$\Lambda^+_c(2940)$ are shown in Fig.~\ref{Fig:tcsplus}. The dashed,
dotted, and dash-dotted curves stand for the contributions from the
$s$ channel, $t$ channel, and $u$ channel, respectively. Their total
contribution is shown by the solid line. In Fig.~\ref{Fig:tcsplus},
the blue line stands for the contributions from the ground
$\Lambda^+_c(2286)$ state. One can see that the $t$-channel $D^{*0}$
meson exchange plays a predominant role, while contributions from
the $s$ channel nucleon pole, and $u$ channel $\Sigma_c$ exchange
are small. The dominant $D^{0*}$ exchange contribution can be easily
understood since the $\Lambda^+_c(2940)$ resonance is assumed as a
molecular state of $D^{*0} p$. In addition, the contribution from
$\Lambda^+_c(2286)$ is also important especially for the very close
to threshold region. Besides, there is no contributions from $D$
meson exchange in the $t$ channel. Hence, this reaction provides a
good platform for studying the $\Lambda^+_c(2940)$ resonance with
the assumption that it is a molecular $D^{*0} p$ state.

It is worth mentioning that the numerical results are sensitive to
the value of the cutoff parameter $\Lambda$. To see how much it
depends on the cutoff parameter, we also show by the red solid curve
in Fig.~\ref{Fig:tcsplus} the theoretical result for the total
contributions with $\Lambda = 2.5$ GeV for comparison. We see that
the total cross section reduces by a factor of 10 when $\Lambda$
decreases from $3$ to $2.5$ GeV.

\begin{figure}[htbp]
\begin{center}
\includegraphics[scale=0.45]{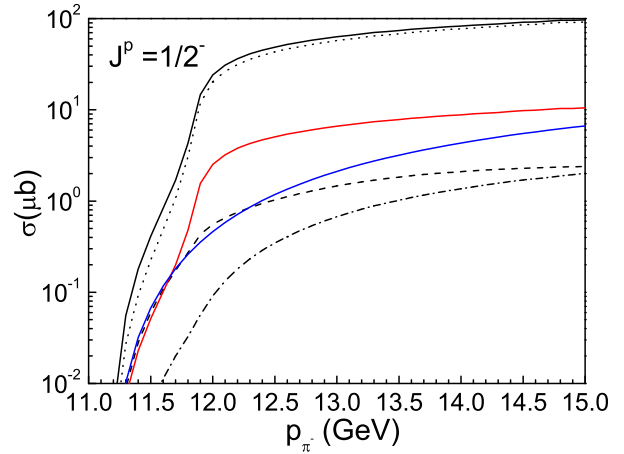}
\caption{(Color online) As shown in Fig.~\ref{Fig:tcsplus} but for
$J^P = \frac{1}{2}^-$ of the $\Lambda^+_c(2940)$.}
\label{Fig:tcsminus}
\end{center}
\end{figure}

The results for $J^P = \frac{1}{2}^-$ of the $\Lambda^+_c(2940)$ are
shown in Fig.~\ref{Fig:tcsminus}. We can see that the total cross
sections are larger than the case of $J^P = \frac{1}{2}^+$, and the
$t$-channel $D^{*0}$ exchange is also predominant. In this case, the
contribution from the ground $\Lambda^+_c(2286)$ state is less
important than in the case of $J^P = \frac{1}{2}^+$ for
$\Lambda^+_c(2940)$ resonance near the threshold region.

From Figs.~\ref{Fig:tcsplus} and \ref{Fig:tcsminus}, we see a clear
sharp growing around $p_{\pi^-} = 12$ GeV which is because at that
energy point, the invariant mass of $D^0 p$ system will reach and
pass by $2.94$ GeV~\footnote{The maximal value of the invariant mass
of the $D^0 p$ system is $\sqrt{s} - m_{D^-}$ with $s = m^2_{\pi^-}
+ m^2_p + 2m_p \sqrt{p^2_{\pi^-} + m^2_{\pi^-}}$ the invariant mass
square of the $\pi^- p$ system. It is easy to get $\sqrt{s} -
m_{D^-} = 2.97$ GeV with $p_{\pi^-} = 12$ GeV.} that is the mass of
the $\Lambda^+_c(2940)$ resonance, the propagator $\frac{1}{q^2 -
M^2 + i M \Gamma}$ of the $\Lambda^+_c(2940)$ resonance will give a
large contribution because of its narrow total decay width.
Furthermore, a change of the spin parity assignment from
$\frac{1}{2}^+$ to $\frac{1}{2}^-$ leads to an enhancement of the
total cross section by a factor of more than 10, as found in
Ref.~\cite{Dong:2014ksa}. However, as discussed before, our
theoretical result on the total cross section of the $\pi^- p \to
D^- D^0 p$ reaction is sensitive to the cutoff $\Lambda$. Thus we
cannot adjust the parity of $\Lambda_c(2940)$ from the total cross
section of the $\pi^- p \to D^- D^0 p$ reaction. We should study
other observables to distinguish the two parity assignments.

\subsection{Differential cross sections}

In addition to the total cross section, we studied also the
invariant mass and angle distributions for the $\pi^- p \to D^- D^0
p$ reaction. Unfortunately, we cannot distinguish the two
spin-parity assignments from those first order differential cross
sections. This is because the $D^0 p$ angular distribution is
determined solely by the spin of $\Lambda^+_c(2940)$ and not its
parity~\cite{Moriya:2014kpv}. Furthermore, the contributions from
$u$ channel and $s$ channel are too small to affect the mass
distributions of $D^0 p$ for the two spin-parity assignments, which
means the mass distributions are almost the same for the two cases.
In order to see the difference between the two assignments of the
$\Lambda_c(2940)$ resonance, we further move to study the second
order differential cross section of the $\pi^- p \to D^- D^0 p$
process.

The second order differential cross section for the process $\pi^- p
\to D^- D^0 p$ is obtained through the expression
\begin{eqnarray}
\frac{d^2\sigma}{dM_{D^0 p}d\Omega} &=& \frac{m^2_p}{2^9 \pi^5
\sqrt{s[(p_1 \cdot p_2)^2 - m^2_{\pi^-} m^2_p]}}  \nonumber \\
&& \times \int \sum_{s_i, s_f} |{\cal M}|^2 |\vec{p}_3|
|\vec{p}^{~*}_5| d\Omega^* ,
\end{eqnarray}
where $|\vec{p}^{~*}_5|$ and $\Omega^*$ are the three-momentum and
solid angle of the outing proton in the center-of-mass (c.m.) frame
of the final $D^0 p$ system, while $|\vec{p}_3|$ and $\Omega$
($\theta, \phi$) are the three-momentum and solid angle of the final
$D^-$ meson in the c.m. frame of the initial $\pi^- p$ system. In
the above equation $M_{D^0 p}$ is the invariant mass of the final
$D^0 p$ two-body system, and $s$ is the invariant mass square of the
$\pi^- p$ system.

\begin{figure}[htbp]
\begin{center}
\includegraphics[scale=0.45]{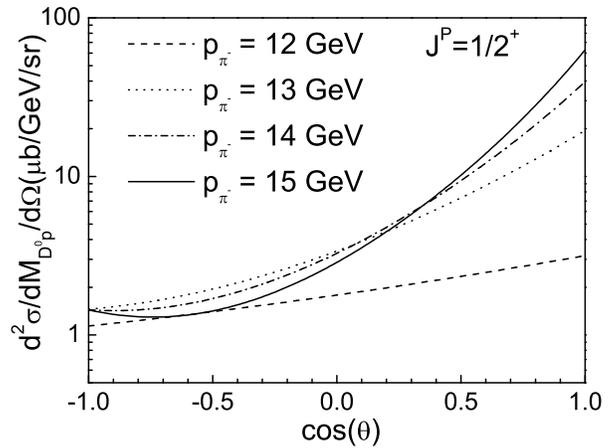}
\caption{Differential cross sections for the $\pi^- p \to D^- D^0 p$
reaction as a function of the scattering angle ($\theta$) of the
outgoing $D^-$ meson in the c.m. frame of $\pi^- p$ system for $J^P
= \frac{1}{2}^+$ of the $\Lambda^+_c(2940)$.} \label{Fig:dcsplus}
\end{center}
\end{figure}

\begin{figure}[htbp]
\begin{center}
\includegraphics[scale=0.45]{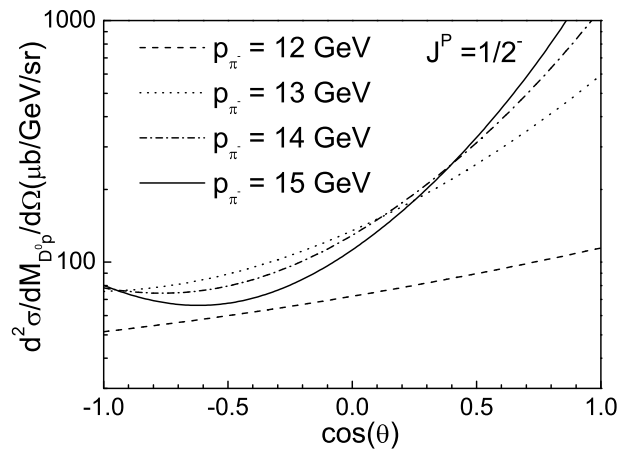}
\caption{As shown in Fig.~\ref{Fig:dcsplus} but for $J^P =
\frac{1}{2}^-$ of the $\Lambda^+_c(2940)$.} \label{Fig:dcsminus}
\end{center}
\end{figure}

The numerical results obtained with $\Lambda = 3$ GeV at $M_{D^0 p}
= 2940$~MeV~\footnote{At this energy point, the contribution from
ground $\Lambda_c^+(2286)$ state will be very small comparing with
$\Lambda^+_c(2940)$ resonance because of the narrow total decay
width of the $\Lambda^+_c(2940)$ resonance.}, for the case of $J^P =
\frac{1}{2}^+$ and $J^P = \frac{1}{2}^-$ for the
$\Lambda^+_c(2940)$, are shown in Figs.~\ref{Fig:dcsplus} and
\ref{Fig:dcsminus}, respectively. In those figures, the dashed,
dotted, dash-dotted, and solid curves stand for the results obtained
at $p_{\pi^-} = 12$, $13$, $14$, and $15$ GeV, respectively. We see
that our theoretical numerical results of the differential cross
sections for the two assignments are different and can be easily
distinguished . Therefore, this observable can be employed, in the
future experiments at J-PARC, to tell the intrinsic parity of the
$\Lambda_c(2940)$ resonance.

\begin{figure}[htbp]
\begin{center}
\includegraphics[scale=0.45]{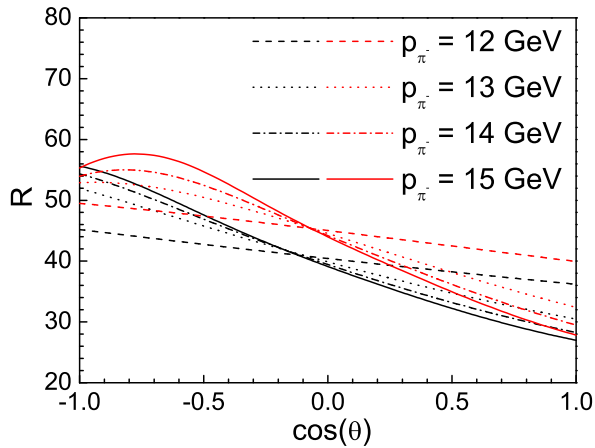}
\caption{(Color online) Ratio of the differential cross sections for
$J^P = \frac{1}{2}^-$ and $J^P = \frac{1}{2}^+$. The black and red
curves are obtained with $\Lambda = 3$ and $2.5$ GeV, respectively.}
\label{Fig:dcsratio}
\end{center}
\end{figure}

To see clearly how different the differential cross sections for the
two assignments, we define the ratio $R$ as
\begin{eqnarray}
R = \frac{\frac{d^2\sigma}{dM_{D^0 p}d\Omega} (J^P =
\frac{1}{2}^-)}{\frac{d^2\sigma}{dM_{D^0 p}d\Omega} (J^P =
\frac{1}{2}^+)},     \label{dcsration}
\end{eqnarray}
which will be not flat vs ${\rm cos}\theta$ if the shape of the
differential cross sections for the two assignments are different.
Furthermore, the ratio $R$ is not sensitive to the value of the
cutoff parameter $\Lambda$. We show the numerical results for $R$ in
Fig.~\ref{Fig:dcsratio} with $\Lambda = 3$ (black curves) and $2.5$
GeV (red curves). We see clearly that $R$ is not flat as a function
of ${\rm cos}\theta$, it changes dramatically. This phenomenon tells
that the shapes of the second order differential cross section
$\frac{d^2\sigma}{dM_{D^0 p}d\Omega}$ for the two assignments $J^P =
\frac{1}{2}^{\pm}$ for the $\Lambda^+_c(2940)$ resonance are sizably
different. We hope that this feature may be used to determine the
parity of the $\Lambda^+_c(2940)$ resonance.

\section{Summary}

In this work, we have studied the $\pi^- p \to D^- D^0 p$ reaction
near threshold within an effective Lagrangian approach. In addition
to the $s$ channel nucleon pole, $u$ channel $\Sigma_c$ exchange,
the $D^{*0}$ meson exchange in the $t$ channel is also investigated
by the assumption that the $\Lambda^+_c(2940)$ is a molecular
$D^{*0} p$ state. The total and differential cross sections are
predicted. Our results show that the $t$-channel $D^{*0}$ exchange
is predominant, and also a change of the spin-parity assignment for
the $\Lambda^+_c(2940)$ resonance from $\frac{1}{2}^+$ to
$\frac{1}{2}^-$ leads to an enhancement of the total cross section
by a factor of more than $10$. Furthermore, it is found that the
theoretical numerical results of the second order differential cross
sections, $d^2\sigma/dM_{D^0 p}/d\Omega$, of the two assignments are
sizably different. This conclusion can be easily distinguished and
may be tested by the future experiments at J-PARC.

\section*{Acknowledgments}

This work is partly supported by the National Natural Science
Foundation of China under Grants No. 11475227, No. 10775148, No.
10975146, No. 11475192, No. 11035006, and No. 11261130, as well as
supported, in part, by the DFG and the NSFC through funds provided
to the Sino-German CRC 110 "Symmetries and the Emergence of
Structure in QCD".

\bibliographystyle{unsrt}

\end{document}